\documentclass[fleqn,usenatbib]{mnras}

\usepackage{newtxtext,newtxmath}

\usepackage[T1]{fontenc}

\DeclareRobustCommand{\VAN}[3]{#2}
\let\VANthebibliography\thebibliography
\def\thebibliography{\DeclareRobustCommand{\VAN}[3]{##3}\VANthebibliography}

\usepackage{graphicx}	
\usepackage{amsmath}	
\usepackage{mathtools}
\usepackage[justification=centering]{caption}
\title[Constraining the Cosmological Constant Using PINNs]{Constraining the Cosmological Constant from Stellar Orbits \\Around Sgr A* Using Physics-Informed Neural Networks}

\author[Eyama \& Masada]{
Shinsei Eyama,$^{1}$\thanks{E-mail: dd250001@g.hit-u.ac.jp}
Youhei Masada,$^{2}$
\\
$^{1}$Department of Social Data Science, Hitotsubashi University, Kunitachi, Tokyo 186-8601, Japan\\
$^{2}$Department of Applied Physics, Faculty of Science, Fukuoka University, Fukuoka 814-0180, Japan
}

\date{Accepted XXX. Received YYY; in original form ZZZ}

\pubyear{\the\year{}}

\begin{document}
\label{firstpage}
\pagerange{\pageref{firstpage}--\pageref{lastpage}}
\maketitle

\begin{abstract} 
We present a novel analytical framework employing Physics-Informed Neural Networks (PINNs) to constrain the cosmological constant $\Lambda$ through the analysis of stellar orbits around the supermassive black hole (SMBH) Sgr A* at the Galactic center. Focusing on the well-observed S2 star, we use an inverse PINN (iPINN) architecture to infer orbital elements and estimate the total precession angle from astrometric data. By isolating the contribution from $\Lambda$, which is defined as the difference between the total precession and the Schwarzschild precession, we derive a stringent upper bound of $\Lambda \leq 5.67 \times 10^{-40}, \mathrm{m}^{-2}$, which is approximately two orders of magnitude tighter than previous estimates obtained using similar data-driven methods. Extension of our analysis to two additional long-period S-stars, S1 and S9, reveals that while the cosmological precession becomes relatively more prominent in such systems, limited orbital coverage introduces significant uncertainties in parameter estimation. Among the cases examined, the constraint derived from S2 remains the most robust. Our results highlight the potential of PINN-based approaches for extracting physical insights from sparse or noisy astronomical data. Future applications to next-generation observational data and further methodological improvements in machine learning are expected to refine the cosmological constraints and enable broader tests of gravitational theories.
\end{abstract}




\section{Introduction}
The term $\Lambda$, known as cosmological constant, was originally introduced to Einstein's field equations of general relativity to achieve a static universe \citep{einstein17}. 
Currently, it is used to explain the accelerating expansion of the universe and is considered the simplest representation of ``dark energy", which is an unknown form of energy known to be the cause of this acceleration. \citep[see, e.g,][and references therein]{padmanabhan03, riess+98, perlmutter+99}.
For understanding the universe and uncovering its future trajectory, quantitatively evaluating the magnitude of $\Lambda$ is of paramount importance \citep[e.g.,][]{bennett13, planck14}.

An intriguing approach to estimating the magnitude of $\Lambda$ involves analyzing the precession of stellar orbital motion. According to Einstein's field equations, spacetime curvature induces the precession of stellar orbits, a phenomenon known as ``Schwarzschild precession" \citep[e.g.,][]{misner+73,will14}. When the $\Lambda$ term is included, an additional effect is introduced -referred to as "cosmological precession", which further influences stellar orbital motion \citep[e.g.,][]{kerr03}. Notably, the cosmological precession exhibits a strong dependence on the semimajor axis of the orbit (see eq. (11)) and becomes significantly more pronounced in regions close to a supermassive black hole (SMBH). Hence, the S-stars orbiting Sgr A$^{\star}$ at the center of our galaxy are promising candidates for measuring the contribution of the $\Lambda$ term to stellar precession \citep[see, e.g.,][for Sgr A$^{\star}$ and the S-stars orbiting it ]{balick+74,Fulvio+01}.

The S-stars are predominantly studied through electromagnetic observations conducted using the Very Large Telescope (VLT) at the European Southern Observatory. Over a span of 25 years, \citet{gillessen17} have meticulously tracked these stars; however, due to their extensive orbital periods, ranging from several decades to centuries, most have yet to complete a full revolution \citep[see][for reviews]{genzel+10}. 
Among them, S2, which is identified as an early B-type main-sequence star with an orbital period of approximately 16 years, is the sole star for which a complete orbit has been observed. In particular, \citet{abuter20} reported that the point of closest approach of S2 to the SMBH shifts by about $12'$ per orbit, which is consistent with the ``Schwarzschild precession" around Sgr A$^{\star}$ with a mass of approximately 4 million solar masses \citep[c.f.,][]{abuter+18a,abuter+18b,abuter+19,do+19}.
See. e.g., \citet{dayem+24} for recent related studies by the GRAVITY Collaboration

The S-stars themselves are subjects of active research regarding their formation mechanisms \citep[e.g.,][]{lockmann+08, figer+09}, and their orbits provide powerful probes for testing properties of the Galactic Center black hole \citep{merritt+10} and even alternative gravity theories, such as those involving scalar fields \citep{bambhaniya+24}. 

The potential impact of $\Lambda$ on the precession of S2 was explored for the first time by \citet{galikyan23}. Utilizing a novel machine learning (ML) framework based on Physics-Informed Neural Networks (PINNs) \citep[e.g.,][]{raissi17i}, they derived an upper limit for $\Lambda$ as
\begin{equation}
\mathmakebox[0.9\columnwidth]{
\Lambda \leq 5.8 \times 10^{-38} \ \ {\rm m}^{-2}\;.
} 
\end{equation}
Building on these findings, in \citet{galikyan24}, they conducted a further analysis of S2's precession, placing an upper bound on the density of the star cluster surrounding Sgr A$^{\star}$. In \S~2, the PINN-based ML method, which is used in \citet{galikyan23} for constraining the $\Lambda$,  is explained in detail. 

Various methods have been employed to place observational constraints on the cosmological constant. The most stringent bounds to date come from cosmological probes such as the Planck 2018 results combined with baryon acoustic oscillation (BAO) measurements, yielding an upper limit of approximately $\Lambda \lesssim 1.2 \times 10^{-52}$~m$^{-2}$ within the $\Lambda$CDM framework \citep{planck18}. Independent constraints have also been derived from gravitational lensing \citep{ishak+08}, galaxy rotation curves \citep{benisty24b}, binary pulsar systems, Solar System dynamics, and local group measurements \citep{benisty24a}. These studies provide complementary upper bounds, typically ranging between $10^{-34}$ and $10^{-52}$~m$^{-2}$ depending on the astrophysical context and methodology. See also e.g., \citet{Wickramasinghe99}, \citet{Kagramanova+06}, \citet{Sereno+06}, \citet{Iorio18} and \citet{Liang+14} for the upper bounds on $\Lambda$ from the Solar System.

This study aims to investigate the orbital precession of the star S2, which orbits the SMBH (Sgr A*) at the Galactic center, in order to place an independent upper limit on the cosmological constant $\Lambda$.
Unlike previous studies based on large-scale structure, gravitational-lensing, or Solar System dynamics, our approach relies solely on stellar orbital dynamics in the strong gravity regime near the Galactic center.
To this end, we employ Physics-Informed Neural Networks (PINNs) to model the orbital motion of S2 by incorporating both Schwarzschild precession and the effects of $\Lambda$. We investigate how the upper bound of $\Lambda$ responds to changes in the hyperparameters that control the relative contribution of the regression loss and the physical loss in the PINNs framework.

Section 2 outlines the physical model used to describe the orbital precession of S2, with particular attention to the contributions from Schwarzscild curvature and the influence of cosmological constant $\Lambda$. Section 3 describes the PINN-based regression model, which incorporates physical constraints, along with the inverse modeling method (iPINNs) employed for parameter estimation. Section 4 applies the PINN-based model to observational data to predict the orbit of S2 and to constrain the cosmological constant $\Lambda$ by comparing the inferred precession with that predicted by general relativity. The same procedure is extended to other S-stars, including S1 and S9, to examine how the orbital period influences the precision of the resulting constraints. Finally, Section 5 summarizes the main findings of our study, and discusses future prospects for improving the precision of the analysis as more observational data become available.

\section{Physical Model of Orbital Precession} 
\begin{figure}
    \begin{center}
    \includegraphics[scale=0.28]{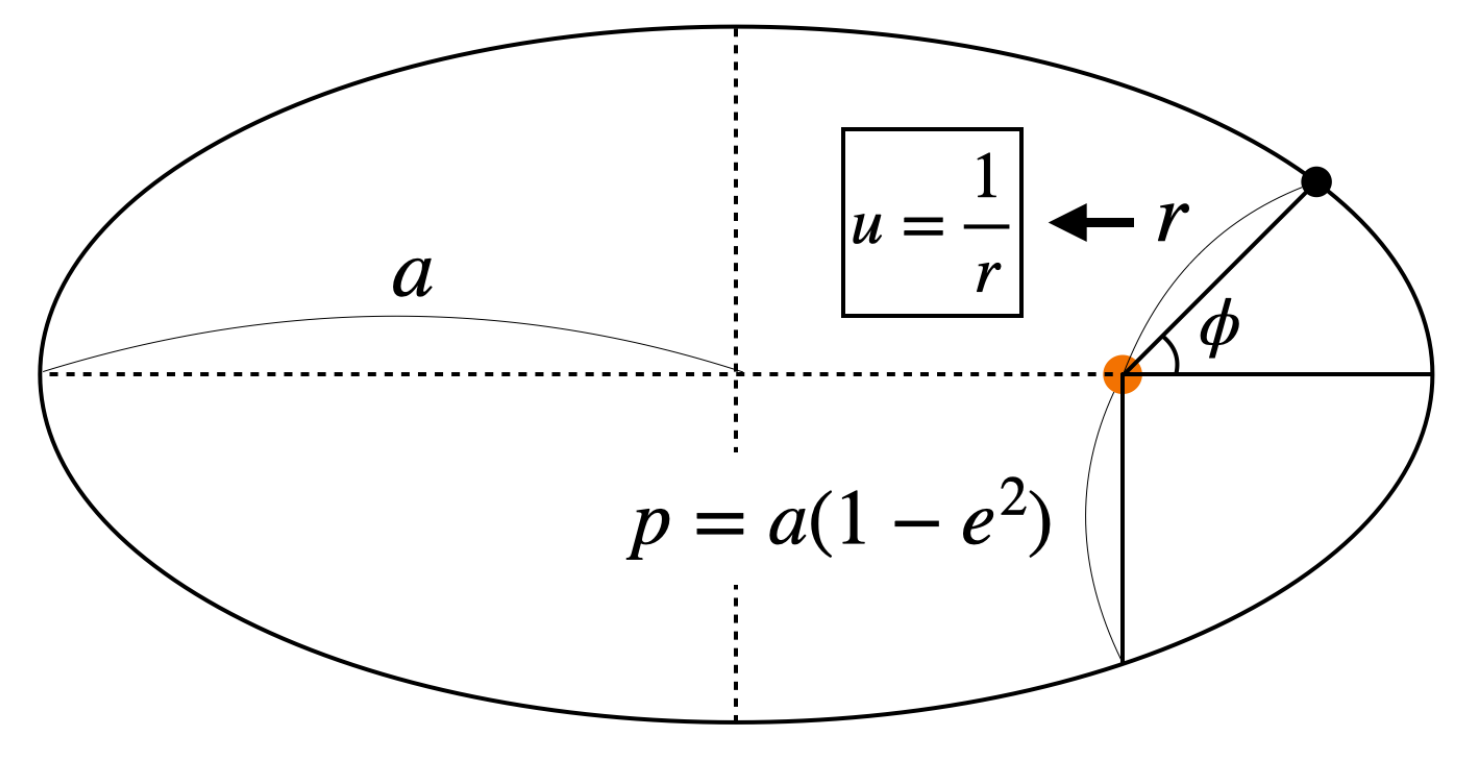}
    \caption{Schematic representation of the orbital geometry. The central mass is located at the intersection of the dashed lines, corresponding to one of the two foci of the ellipse. The radial distance $r$ is measured from the focus to the orbiting object (black dot), and the corresponding inverse radius $u = 1/r$ is the quantity predicted by the PINN model. The angular coordinate $\phi$ denotes the true anomaly, defined as the angle between the periapsis (marked by the orange dot) and the position of the object. The semi-major axis $a$ characterizes the size of the orbit, and the semi-latus rectum $p=a(1-e^2)$ is used to express the orbit equation in terms of $r$ and $\phi$.}
    \label{f1}
    \end{center}
\end{figure}
\subsection{Orbital Geometry}
Fig.\ref{f1} shows schematic representation of the orbital geometry, which is used for our analysis. The orbiting body is shown as a black dot, and the central mass—located at one of the two foci of the ellipse—is represented by the orange dot at the intersection of the dashed lines. The radial distance $r$ is measured from the focus to the orbiting object, and the corresponding inverse radius $u = 1/r$ is the quantity predicted by the PINN model. The angular coordinate $\phi$ denotes the true anomaly, defined as the angle between the periapsis and the position of the orbiting object. The semi-major axis $a$ characterizes the size of the orbit, and the semi-latus rectum defined by 
\begin{equation}
\mathmakebox[0.9\columnwidth]{
p=a(1-e^2) \;, 
}
\end{equation}
is used to express the orbit equation in terms of $r$ and $\phi$. 
\subsection{Schwarzschild Precession} 
The Schwarzschild precession is the relativistic shift in the orbit of an object caused by spacetime curvature around a massive body, as predicted by general relativity. The equation of elliptical motion is given, with eccentricity $e$, declination $\phi$, and semi-major axis $a$, by
\begin{equation}
\mathmakebox[0.9\columnwidth]{
r = \frac{a(1-e^2)}{1+e\,{\rm cos}\phi} \;,
}
\end{equation}
where $r$ is the radial distance measured from the focus to the orbiting object. With the semi-latus rectum $p$ and the inverse of the radius $u (= 1/r)$, eq.(3) can be rewritten as 
\begin{equation}
\mathmakebox[0.9\columnwidth]{
u = \frac{1}{p}(1 + e\cos \phi) \;, 
}
\end{equation}
Introducing the Darwin variable $\chi$, following \citet{chandrasekhar}, we model the orbit in the framework of the Schwarzschild metric as
\begin{gather}
\label{sch0}
\mathmakebox[0.9\columnwidth]{
u = \frac{\mu}{\widetilde{M}}(1 + e\cos \chi) \;, 
}\\
\label{sch1}
\mathmakebox[0.9\columnwidth]{
\left(\frac{d\chi}{d\phi}\right)^2 = 1 - 2\mu(3 + e\cos\chi) \;, 
}\\
\label{sch2}
\mathmakebox[0.9\columnwidth]{
\frac{d^2\chi}{d\phi^2} = \mu e\sin \chi \;,
}
\end{gather}
where $\mu=\widetilde{M}/p$, $\widetilde{M} = GM/c^2$. Here, $M=4.28\times10^6\; M_\odot$ is adopted for the mass of the central body \citep{gillessen17}.
Eq.(\ref{sch1}) can be approximated by a first-order Taylor expansion, with respect to $\mu$, as 
\begin{gather}
\mathmakebox[0.9\columnwidth]{
d\phi \simeq d\chi(1 + 3\mu + \mu e\,{\rm cos}\chi)
\;,} \nonumber \\
\mathmakebox[0.9\columnwidth]{ \leftrightarrow 
\phi = (1 + 3\mu)\chi + \mu e\,{\rm sin}\chi + {\rm const}\;.
}
\end{gather}
The Schwarzscild precession per orbit $\delta\phi_{\rm SP}$ is then derived as 
\begin{eqnarray}
\delta\phi_{\rm SP} &\equiv& \phi_{(\chi=2\pi)} - \phi_{(\chi=0)} - 2\pi \nonumber \\
&=& 6\pi\mu \label{phisp}
\end{eqnarray}
with $\mu = \widetilde{M}/p$. Since the Schwarzschild precession $\delta\phi_{\rm SP}$ depends on the mass of the central body and the semi-latus rectum of the orbit, it becomes particularly detectable in strong gravitational fields. This effect is especially pronounced in short-period stars such as S2, where it can be clearly distinguished from the precession caused by the cosmological constant $\Lambda$.

\begin{figure*}
    \begin{center}
    \includegraphics[scale=0.45]{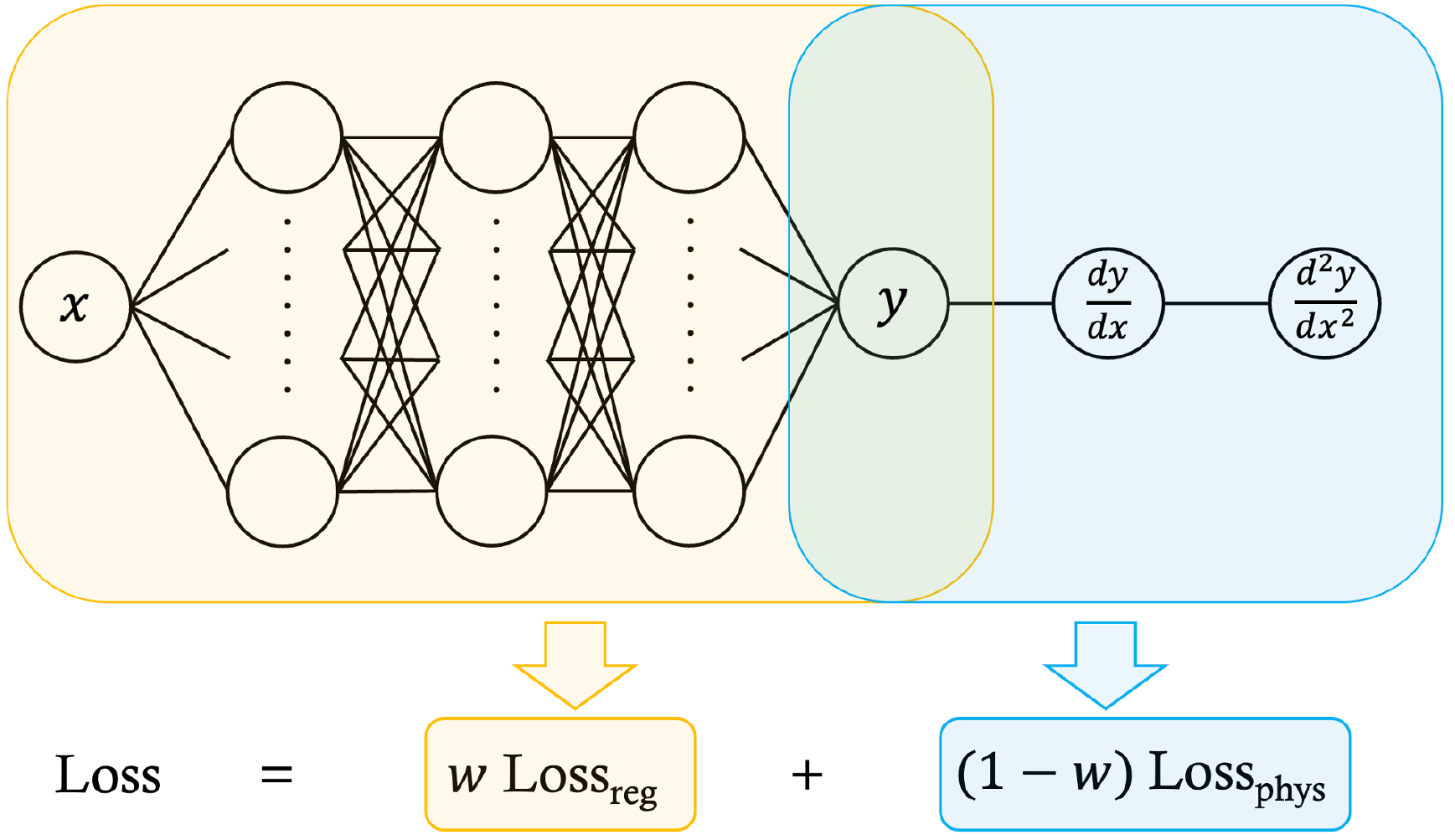}
    \caption{Architecture of the PINN used in this study. The input to the network, i.e., $x$, is the angular coordinate $\phi$, while the output, i.e., $y$, is the predicted inverse radial distance $\hat{u}(\phi) = 1/\hat{r}(\phi)$ in this work. The training process simultaneously minimizes two loss terms: the regression loss ${\rm Loss}_{\rm Reg}$, which quantifies the discrepancy between predicted and observed data points $(u_i,\phi_i)$, and the physical loss ${\rm Loss}_{\rm Phys}$, which enforces consistency with the governing differential equation of motion. The total loss is defined as a weighted sum of these components. By embedding physical knowledge into the loss function, the PINN is capable of producing physically consistent predictions even under noisy or incomplete data conditions.}
    \label{f2}
    \end{center}
\end{figure*}
\subsection{Contribution of the Cosmological Precession due to \texorpdfstring{$\Lambda$}{Lambda} term}
In this section, we summarize the contribution of the cosmological constant $\Lambda$, so-called "cosmological precession" $\delta\phi_{\rm \Lambda}$, to the total orbital precession. Although the $\Lambda$ term induces only a small perturbation to the spacetime geometry, it leads to a minor but non-negligible correction to the orbital precession. The term $\delta\phi_{\rm \Lambda}$ becomes particularly relevant for long-period stars, where the dominant Schwarzschild contribution is suppressed, thereby enhancing the relative significance of the $\Lambda$-induced effect from an observational perspective.

If the total orbital precession inferred from observational data includes both the Schwarzschild component $\delta\phi_{\rm SP}$ and the cosmological correction $\delta\phi_{\Lambda}$, the combined effect can be written as
\begin{equation}
\mathmakebox[0.9\columnwidth]{
\delta\phi_{\rm Reg} = \delta\phi_{\rm SP} + \delta\phi_{\Lambda}\;,
}
\label{deltaphi}
\end{equation}
where $\delta\phi_{\rm Reg}$ represents the total precession obtained from the regression-based orbital solutions predicted by the PINNs. This quantity is computed using the reciprocal $u = 1/r$ of the radial coordinate from Sgr A*, by evaluating the angular shift of the minimum point of $u$ over two orbital periods:
\begin{equation}
\mathmakebox[0.9\columnwidth]{
\delta\phi_{\rm Reg} = \phi_{(\hat{u}_{\rm min1})} - \phi_{(\hat{u}_{\rm min0})} - 2\pi\;.
}
\label{phireg}
\end{equation}
The cosmological precession $\delta\phi_{\Lambda}$ is given by the analytical formula derived by \citet{kerr03}:
\begin{equation}
\mathmakebox[0.9\columnwidth]{
\delta\phi_{\Lambda} = \frac{2a^3(1-e^2)^{\frac{1}{2}}\pi}{r_s}\Lambda\;,
}
\label{philam}
\end{equation}
which enables the estimation of $\Lambda$ through comparison with the regression-based results.

\section{Method \& Regression Model} 
PINNs extend traditional neural networks (NNs) by integrating both regressive loss and physical loss, the latter derived from the governing equations of a physical system \citep{raissi17i,raissi17ii}. This integration enables PINNs to outperform conventional data-driven regression methods by embedding physical constraints into the learning process. Consequently, the predictions produced by PINNs are not only more precise but also inherently consistent with the underlying physics.

PINNs thus offer a powerful and versatile framework for tackling complex physical problems, seamlessly integrating data-driven methods with the rigor of fundamental physical laws. Their flexibility has led to widespread adoption in numerous fields, from particle physics to astrophysics \citep[e.g.,][]{schwartz21,su23,ferrer24}, underscoring their transformative potential in a broad range of scientific disciplines.

In this section, we describe the analysis method used in this study based on PINNs.
\subsection{Transformation of Observational Coordinates} 
In this study, the astrometric data used as training inputs for the PINN are expressed in ``right ascension" and ``declination", measured in milliarcseconds, i.e., (RA [mas], Dec [mas]) \citep{gillessen17}. These celestial coordinates are converted into Cartesian coordinates ($x$ [au], $y$ [au]) on the orbital plane, using the Thiele–Innes constants $A$, $B$, $F$, and $G$, which are functions of the argument of periapsis $\omega$, inclination $i$, and the longitude of the ascending node $\Omega$  \citep{becerra20}  
\begin{equation}
\mathmakebox[0.9\columnwidth]{
\begin{cases}
A = \cos\Omega\:\cos\omega - \sin\Omega\sin\omega\cos i\;,\\
B = \sin\Omega\cos\omega + \cos\Omega\sin\omega\cos i\;,\\
F = -\cos\Omega\sin\omega - \sin\Omega\cos\omega\cos i\;,\\
G = -\sin\Omega\sin\omega + \cos\Omega\cos\omega\cos i\;.
\end{cases}
}
\end{equation}
The transformation from celestial to Cartesian coordinates is then given by 
\begin{equation}
\mathmakebox[0.9\columnwidth]{
x = \frac{FX-GY}{BF-AG}\:,\ \ \ \:y = \frac{AX-BY}{AG-BF}\;,
}
\end{equation}
where $X = {\rm RA}$, $Y = {\rm Dec}$. The resulting Cartesian coordinates ($x_i$, $y_i$) are then used as training data in the PINN. At each training point, the input variable is the angular coordinate
\begin{equation}
\mathmakebox[0.9\columnwidth]{
\phi_i = \arctan \left( \frac{y_i}{x_i} \right) \;, 
}
\end{equation}
and the target variable for computing the MSE is the inverse radial distance
\begin{equation}
\mathmakebox[0.9\columnwidth]{
u_i = \frac{1}{r_i} = \frac{1}{\sqrt{{x_i}^2 + {y_i}^2}}\;.
}
\end{equation}

\subsection{Neural Network Model} 
Fig.\ref{f2} shows the framework of PINNs. The primary distinction between conventional NNs and PINNs lies in the formulation of the loss function. In standard NNs, given an input $x$, output $y$, and corresponding training data $\hat{y}$, the regression loss is typically defined as
\begin{equation}
\label{loss_reg}
\mathmakebox[0.9\columnwidth]{
{\rm Loss_{\rm Reg}} = \frac{1}{N}\sum_{i=1}^{N}(\hat{y}_i-y_i)^2  \;.
}
\end{equation}
This regression loss, ${\rm Loss}_{\rm Reg}$, is used exclusively as the objective function in conventional NNs. In contrast, PINNs incorporate physical knowledge by additionally enforcing the governing differential equation, which can be expressed as
\begin{equation}
\label{Function}
\mathmakebox[0.9\columnwidth]{
F(x, \hat{y}_x, \hat{y}'_x, \hat{y}''_x, ... , \hat{y}^{(n)}_x) = 0\;,
}
\end{equation}
the corresponding physics-based loss is then defined as
\begin{equation}
\label{loss_Phys}
\mathmakebox[0.9\columnwidth]{
{\rm Loss}_{\rm Phys} = \frac{1}{N}\sum_{i=1}^{N}F^2(x, \hat{y}_x, \hat{y}'_x, \hat{y}''_x, ... , \hat{y}^{(n)}_x)\;.
}
\end{equation}
By combining both the data-driven and physics-based components, the overall loss function employed in PINNs takes the form
\begin{equation}
\label{loss_func}
\mathmakebox[0.9\columnwidth]{
{\rm Loss} = w\;{\rm Loss}_{\rm Reg} + (1-w)\;{\rm Loss}_{\rm Phys}\;, 
}
\end{equation}
where $w \in [0,1]$ is a tunable hyperparameter that controls the trade-off between empirical regression accuracy and adherence to physical laws. When $w = 0.5$, both terms are equally weighted.

This model consists of two parts: a regression part and a physical part, which together form the loss function. The specific structure of each component and its respective role are described below.

\underline{\bf Regression component : }
This component computes the mean squared error (MSE) between the observed data and the predicted outputs. It consists of a fully connected neural network composed of an input layer, a hidden layer with a configuration of [16, 32, 16] units, and an output layer. The hyperbolic tangent ($\tanh$) function is used as the activation function in the hidden layers. In this setup, the declination of the elliptical orbit at the observational data points, denoted by $\phi$, is used as the input, while the inverse of the distance from Sgr A *, $u$ serves as the output. The MSE between the predicted value $\hat{u}$ and the observed value $u$ defines the regression loss ${\rm Loss}_{\rm Reg}$.

\underline{\bf Physics-based component : } 
This component shares the same network architecture as the regression part but uses a continuous range of input values $\phi \in [0,4\pi]$ as collocation points. The outputs of the network are subjected to automatic differentiation using PyTorch to obtain the necessary derivatives. The physical loss, ${\rm Loss}_{\rm Phys}$, is then evaluated by computing the residuals of the governing differential equation.

The regression and physical loss terms are weighted and aggregated to form the total loss function. To enhance model performance in the later stages of training, the Adam optimizer is employed in combination with PyTorch’s learning rate scheduler for adaptive learning rate control.

\subsection{Inference of Orbital Parameters via Inverse PINNs}
One of the notable advantages of PINNs is their ability to perform inverse analysis - commonly referred to as inverse PINNs (iPINNs). Although conventional NNs employ a regression loss function to minimize the discrepancy between predicted outputs and training data, PINNs augment this loss with residuals from the governing physical equations. This formulation enables the estimation of unknown parameters by simultaneously incorporating observational data and physical constraints, even when the underlying equations contain undetermined quantities.

In iPINNs, initial guesses for unknown parameters are manually specified, and the training process optimizes both the network weights and the parameter values to minimize the combined loss function. In this study, we treat the eccentricity $e$ and the semi-latus rectum $p$ of the elliptical orbit as unknown parameters in the physical model described by Eqs.~(\ref{sch0})–(\ref{sch2}), which govern the precessional motion of the S2 star. Estimating these parameters enables the model to better fit the observed orbital data and allows for more accurate predictions of the precessional behavior. 

Although both the eccentricity $e$ and the semi-major axis $a$ (which defines $p=a(1-e)^2$) are available from existing observational databases, we intentionally treat them as unknowns to be inferred. This strategy is designed to assess the internal consistency and accuracy of the PINN framework by re-estimating these parameters from a model constrained by physical laws rather than relying exclusively on previously measured values.

Through this process, physically meaningful parameters are automatically optimized as part of the training, rather than being externally imposed. The use of iPINNs thus enables more robust and reliable inference of $e$ and $p$, while mitigating the influence of observational noise and systematic uncertainties. Agreement between the inferred values and those derived from observations provides strong evidence for the validity and consistency of the physical model.
\section{Results \& Discussions}
\begin{figure}
    \includegraphics[scale=0.3]{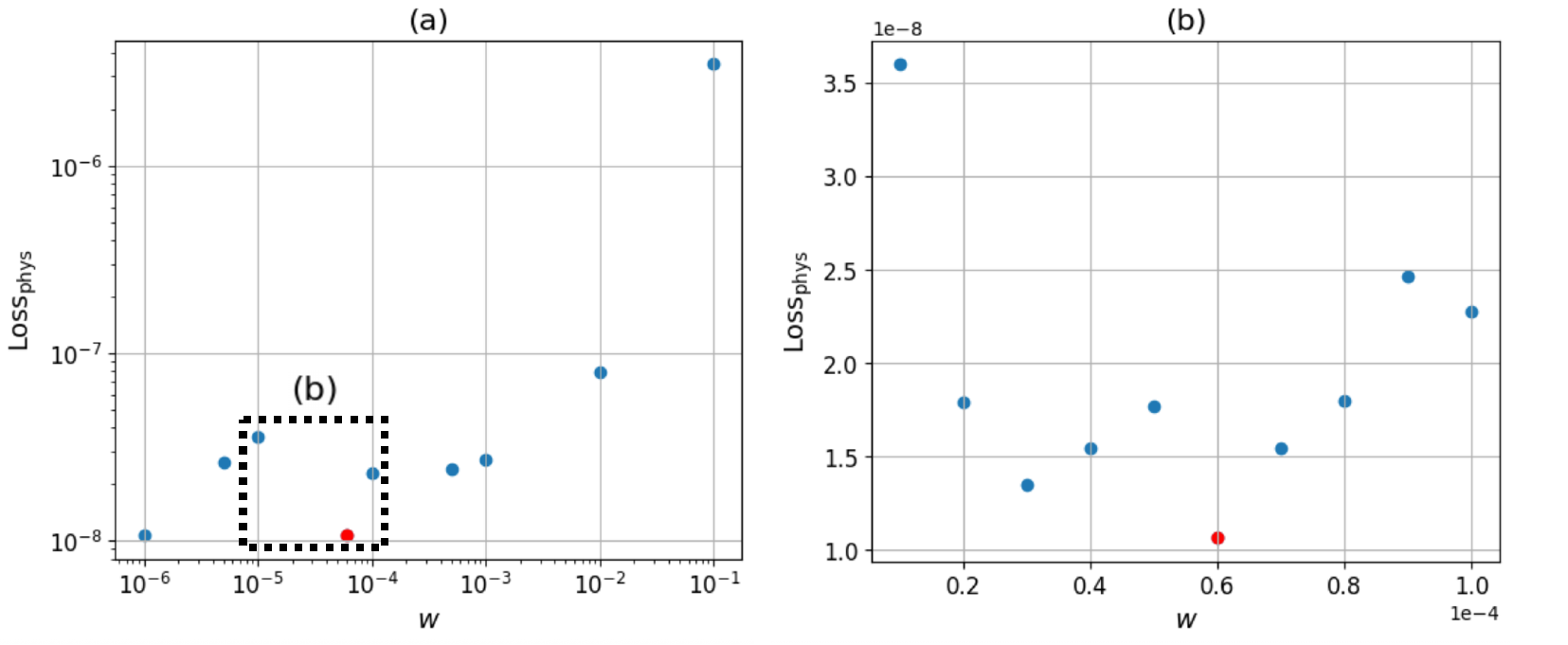}
    \caption{Response of ${\rm Loss}_{\rm phys}$ to variations in the weight parameter $w$. (a) Behavior of ${\rm Loss}_{\rm phys}$ over a broad range of $w$, spanning from $10^{-6} \lesssim w \lesssim 10^{-1} $. (b) Zoomed-in view of the region $10^{-5} \lesssim w \lesssim 10^{-4} $, highlighting the detailed structure of the response. The red dot indicates the value of $w$  yielding the smallest ${\rm Loss}_{\rm phys}$ within the range examined. }
    \label{f3}
\end{figure}

\begin{figure}
    \includegraphics[scale=0.3]{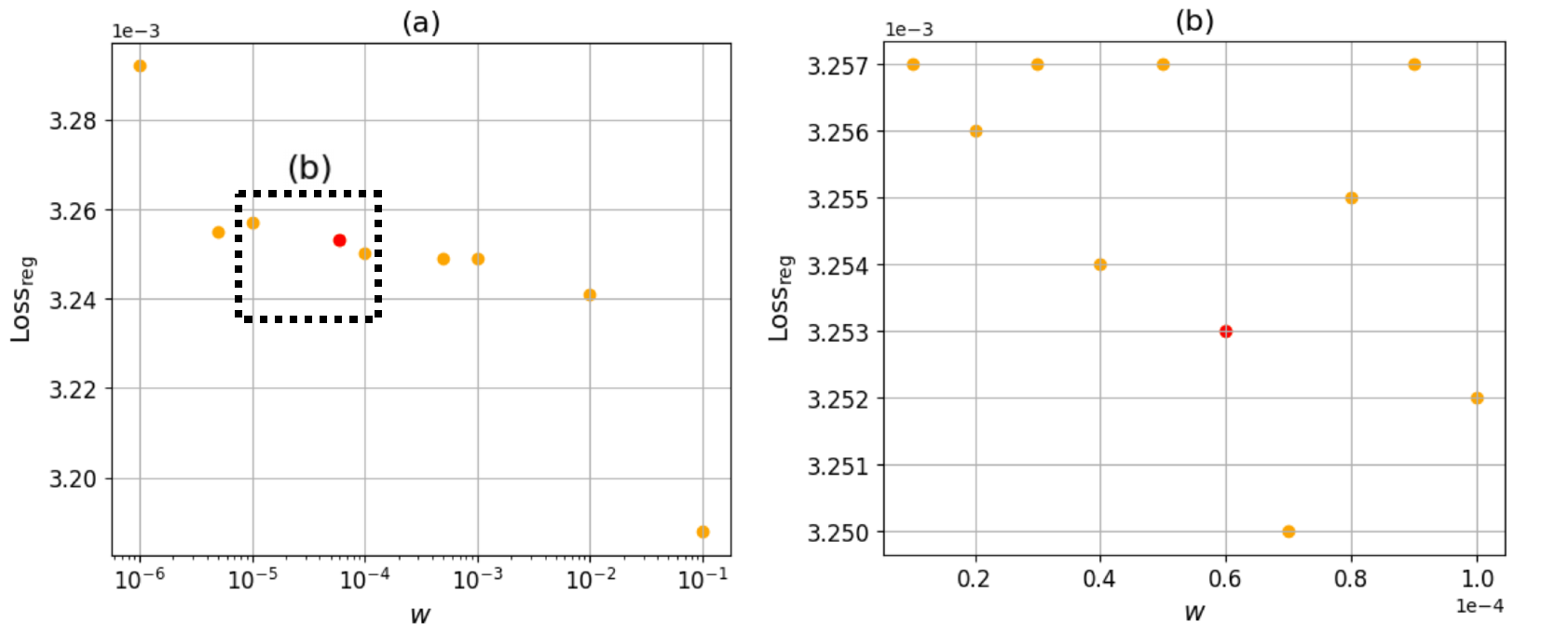}
    \caption{Similar to Fig.3 but for ${\rm Loss}_{\rm Reg}$. }
    \label{f4}
\end{figure}
\begin{figure*}
    \includegraphics[scale=0.5]{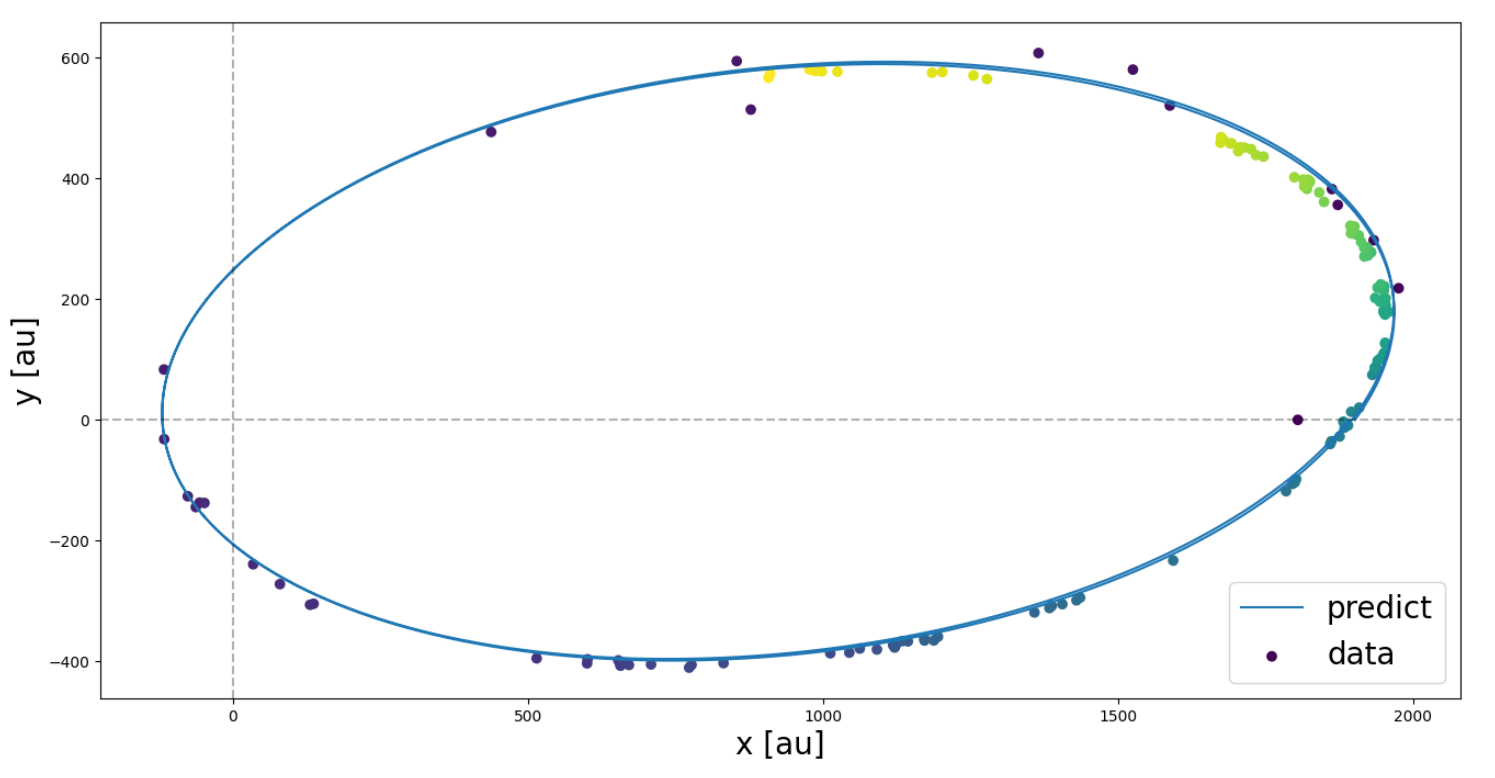}
    \caption{Predicted orbital trajectory of the S2 star, reconstructed using PINNs. The blue solid curve represents the orbit predicted by the PINN, while the colored dots correspond to the observed astrometric data used for training. The color of the data points indicates the phase along the orbit, transitioning from early (dark) to late (bright) stages. 
    The intersection of the dashed lines indicates the position of the central mass.}
    \label{f5}
\end{figure*}
\begin{table*}
    \centering
    \caption{Orbital parameters for S2 inferred via iPINNs. $e$ and $p$ are the orbital parameters for the orbit of the S2 as estimated by \citet{gillessen17}, while $\hat{e}$ and $\hat{p}$ are those inferred via iPINNs.}
    \begin{tabular}{|c|c|c|c|c|c|c|c|}\hline
    $e$ & $\hat{e}$ & $p$ [au] & $\hat{p}$ [au] & $\delta\phi_{\rm Reg}$ ['] & $\sigma_{\rm Reg}$ ['] & $\delta\phi_{\rm SP}$ ['] & $\sigma_{\rm SP}$ [']\\ \hline
    0.884 & 0.8861 & 228 & 225.25 & 12.1491 & $4.1\times10^{-4}$ & 12.1485 & $1.6\times10^{-9}$ \\ \hline
    \end{tabular}
    \label{table1}
\end{table*}
\subsection{Optimal Selection of the Loss Weight \texorpdfstring{$w$}{w}}
A key factor in determining the upper limit of $\Lambda$ is the weight parameter $w$, which governs the trade-off between ${\rm Loss}_{\rm Reg}$ and ${\rm Loss}_{\rm phys}$ in the loss function of PINNs. As defined in Eq.~(\ref{loss_func}), $w$ is a hyperparameter ranging from $0$ to $1$ that determines the relative contribution of each loss term during training. Since ${\rm Loss}_{\rm Reg}$ and ${\rm Loss}_{\rm phys}$ typically operate on different numerical scales, adequate weighting is essential to ensure a balanced learning process. If $w$ is not chosen carefully, the model may either overfit the observational data while neglecting the physical constraints, or conversely, enforce the governing equations too strongly at the cost of observational accuracy.

To identify an appropriate value for $w$, we systematically analyze the behavior of each loss term as a function of $w$. Each loss term is normalized using its value in the first epoch, and manual weighting is applied. Since the magnitude of each loss term varies with $w$, it is crucial to evaluate their sensitivity to this to find the optimal balance.

Figure~\ref{f3}(a) shows the response of ${\rm Loss}_{\rm phys}$ to variations in the weight parameter $w$ over a broad range. Each loss is evaluated after training has reached a steady state where convergence is sufficiently achieved. This coarse survey reveals that the physical loss attains a minimum within the interval $10^{-5} \le w \le 10^{-4}$. To explore this region in greater detail, Figure~\ref{f3}(b) presents the result of a finer search, where 10 equally spaced values of $w$ are sampled within the interval. This detailed analysis identifies $w = 6.0 \times 10^{-5}$ as the value that minimizes ${\rm Loss}_{\rm phys}$, indicating an optimal trade-off between the regression accuracy and physical consistency for this study. 

Although ${\rm Loss}_{\rm phys}$ continues to decrease slightly beyond $w = 10^{-5}$, this improvement is accompanied by a noticeable increase in ${\rm Loss}_{\rm reg}$, reflecting the reduced fidelity to the fitting of observational data. To clarify this trade-off, Figure~\ref{f4} complements Figure~\ref{f3} by showing the corresponding response of ${\rm Loss}_{\rm reg}$ to variations in the weight parameter $w$ over the same range.
These contrasting trends underscore the need to select $w$ so that ${\rm Loss}_{\rm phys}$ is minimized without a significant penalty in regression accuracy. Based on this result, we adopt $w = 6.0 \times 10^{-5}$ in the following analysis, as it provides the optimal balance between physical and observational consistency.

These results underscore the importance of conducting a broad parameter search for $w$ in order to enhance the accuracy of the estimated upper bound of $\Lambda$, while maintaining a proper balance between physical and observational consistency. Notably, previous studies have not examined the optimal choice of $w$ in detail, leaving its impact on physical inference largely unexplored. In this study, we demonstrate that a careful tuning of $w$ not only improves model consistency but also leads to significantly tighter constraints on $\Lambda$, highlighting a key contribution of our approach.

\subsection{Predicted Orbit of S2 with PINNs} 
Figure~\ref{f5} shows the predicted orbital trajectory of the S2 star, reconstructed using PINNs. The horizontal and vertical axes show the Cartesian coordinates $x$ and $y$ on the orbital plane in astronomical units (a.u.). The blue curve represents the orbit predicted by the PINN, while the colored dots correspond to the observed astrometric data used for training. The color of the data points indicates the phase along the orbit, transitioning from early (dark) to late (bright) stages.

Despite the presence of noise and uneven sampling in the observational data—particularly in regions where the orbital motion is rapid—the PINN successfully reproduces a smooth and physically consistent trajectory. Notably, the model captures not only the overall orbital shape but also a subtle shift in the orbit’s orientation, reflecting the relativistic precession of the periapsis. Although the precession is small compared to the full orbit, the deviation is detectable in the predicted trajectory, demonstrating the model’s sensitivity to fine dynamical effects.

This result highlights the capability of PINNs to infer hidden physical features, such as periapsis precession, by enforcing consistency with the governing equations of motion. Such precision would be difficult to achieve using conventional regression techniques alone, particularly under limited or noisy data conditions.

In addition to accurately reconstructing the orbital trajectory, the iPINN framework also provides reliable estimates of the orbital parameters. As summarized in Table~\ref{table1}, the inferred values of eccentricity $\hat{e}$ and semi-latus rectum $\hat{p}$ show excellent agreement with the observational values reported by \citet{gillessen17}. The deviation is within $0.3\%$, underscoring the model’s ability to recover physically meaningful parameters. This consistency further validates the effectiveness of our approach in constraining orbital dynamics and, ultimately, the cosmological constant $\Lambda$.
\begin{figure}
    \includegraphics[scale=0.5]{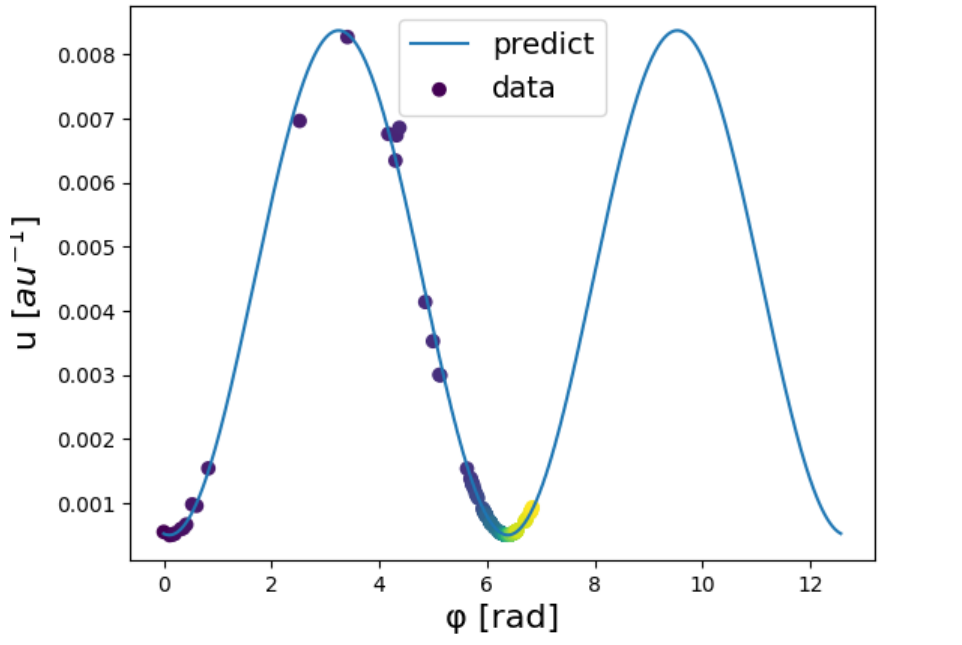}
    \caption{Prediction of inverse radius $u = 1/r$ for the S2 orbit as a function of phase angle $\phi$. The blue curve shows the predicted trajectory from the trained PINN, while the dots represent the observed data. The prediction is extended over the interval $0 \le \phi \le 4\pi$, and is used to evaluate the total precession angle $\delta \phi_{\rm Reg}$ relevant for constraining $\Lambda$.}
    \label{f6}
\end{figure}

Figure~\ref{f6} shows the predicted inverse radial distance $u = 1/r$ as a function of the orbital phase angle $\phi$. The blue curve represents the output of the trained PINN model, while the black-to-yellow dots correspond to the observational data used in training. The prediction is extended over a full phase range from $0$ to $4\pi$, covering two full orbital periods. This extended prediction is essential for identifying the angular positions of the periapsis passage and evaluating the total precessional shift $\delta \phi_{\rm Reg}$, which is subsequently used to constrain the cosmological constant $\Lambda$.

The values of $\delta\phi_{\rm SP}$ and $\delta\phi_{\rm Reg}$ are computed using Eq.(\ref{phisp}) and (\ref{phireg}), respectively. The cosmological precession $\delta\phi_{\Lambda}$ is then calculated as the difference $\delta\phi_{\rm Reg} - \delta\phi_{\rm SP}$, in accordance with Eq.(\ref{deltaphi}). To estimate the upper bound on the cosmological constant $\Lambda$ using Eq.(\ref{philam}), the variances associated with $\delta\phi_{\rm Reg}$ and $\delta\phi_{\rm SP}$ are taken into account. Specifically, the upper bound of $\Lambda$ is determined using the extremal combination $\delta\phi_{\rm Reg}(+3\sigma_{\rm Reg})$ and $\delta\phi_{\rm SP}(-3\sigma_{\rm SP})$, which yields the maximum plausible value of $\delta\phi_{\Lambda}$. This use of the $3\sigma$ range is justified by the fact that, under the assumption of a normal distribution, approximately  $99.7\%$ of the data lie within $\pm3\sigma$ of the mean. 

The upper bound on the cosmological constant $\Lambda$ derived from the results is given by
\begin{equation}
\mathmakebox[0.9\columnwidth]{
\Lambda \leq 5.67 \times 10^{-40}\, \ {\rm m}^{-2}\;.
}
\end{equation}
The application of PINNs allows for the reconstruction of physically consistent orbital trajectories, even in the presence of noise and incompleteness in the observational data. This enables highly accurate predictions that are often difficult to achieve using conventional methods. Compared to the results of a previous study by \citet{galikyan23}, which employed a similar approach, the constraint obtained in this study is approximately two orders of magnitude tighter.

While relatively coarse sampling is sufficient for reconstructing the overall orbital shape, resolving the small angular shift due to relativistic precession requires much finer resolution. To improve precision, the sampling interval in the phase coordinate $phi$ should ideally be smaller than the target precession angle $\delta\phi_{\Lambda} (\simeq 10^{-5} \ {\rm rad})$. Although reducing the interval enhances resolution, it also increases the number of evaluation points and thus the computational cost. Accordingly, when evaluating the full orbit (see Fig.~\ref{f6}), we initially employed a relatively coarse sampling. After identifying the angular locations of the periapsis passages, $u_{\rm min0}$ and $u_{\rm min1}$, the sampling density was locally increased around these regions. This adaptive strategy enabled precise detection of the precessional shift while keeping the overall computational burden manageable.

\subsection{Extension to Other Stars: S1 and S9} 
\begin{figure*}
    \includegraphics[scale=0.55]{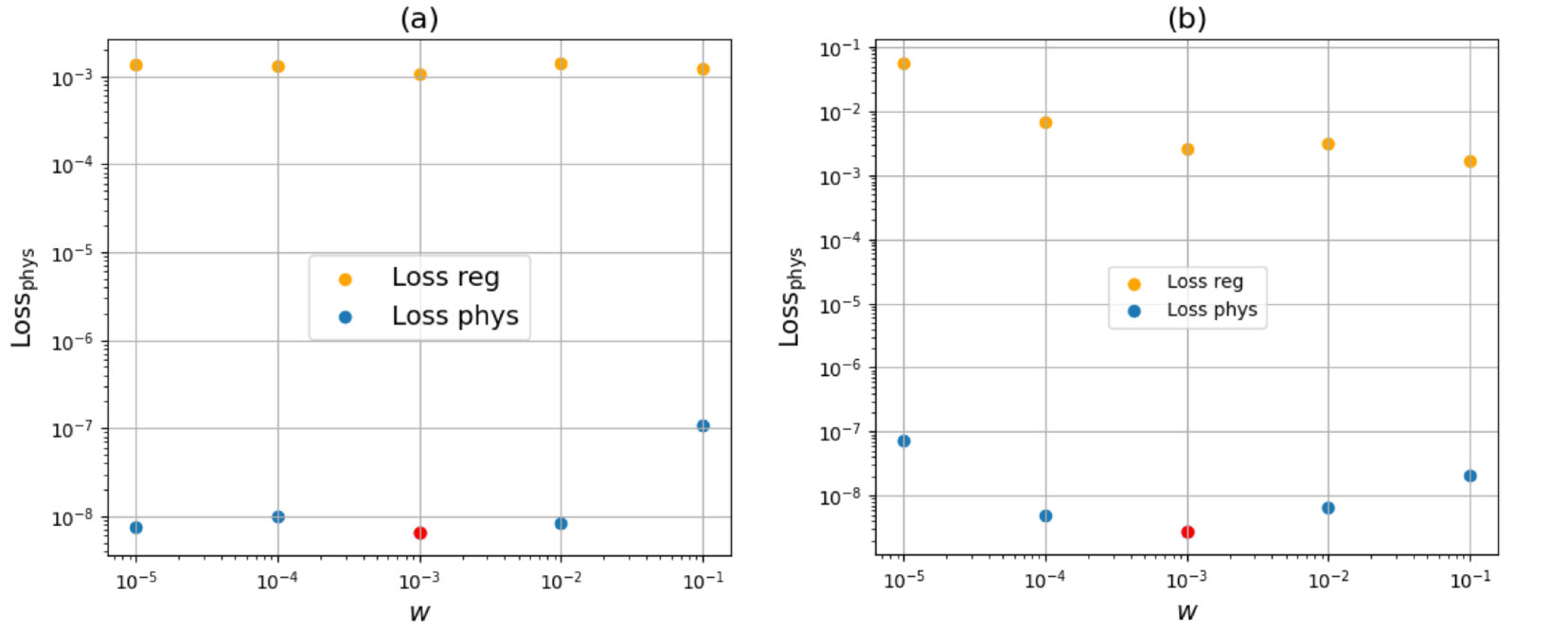}
    \caption{Loss responses as a function of the weight parameter $w$ for (a) S1 and (b) S9. The red dot marks the $w$ value that yields the minimum ${\rm Loss}_{\rm Phys}$ and is used in the subsequent analysis.}
    \label{f7}
\end{figure*}
To assess the generality of our approach and examine whether tighter constraints on $\Lambda$ can be obtained from stars other than S2, we applied the same iPINN-based analysis to two additional members of the S-star cluster: S1 and S9. 

Shown in Figure~\ref{f7} is the loss responses as functions of the weight parameter $w$ for (a) S1 and (b) S9. The yellow and blue dots represent ${\rm Loss}_{\rm Reg}$ and ${\rm Loss}_{\rm Phys}$, respectively. The red dot marks the optimal value $w$ minimizing the physical loss, which is adopted for further analysis in each case. 

\begin{figure}
\centering
\includegraphics[scale=0.4]{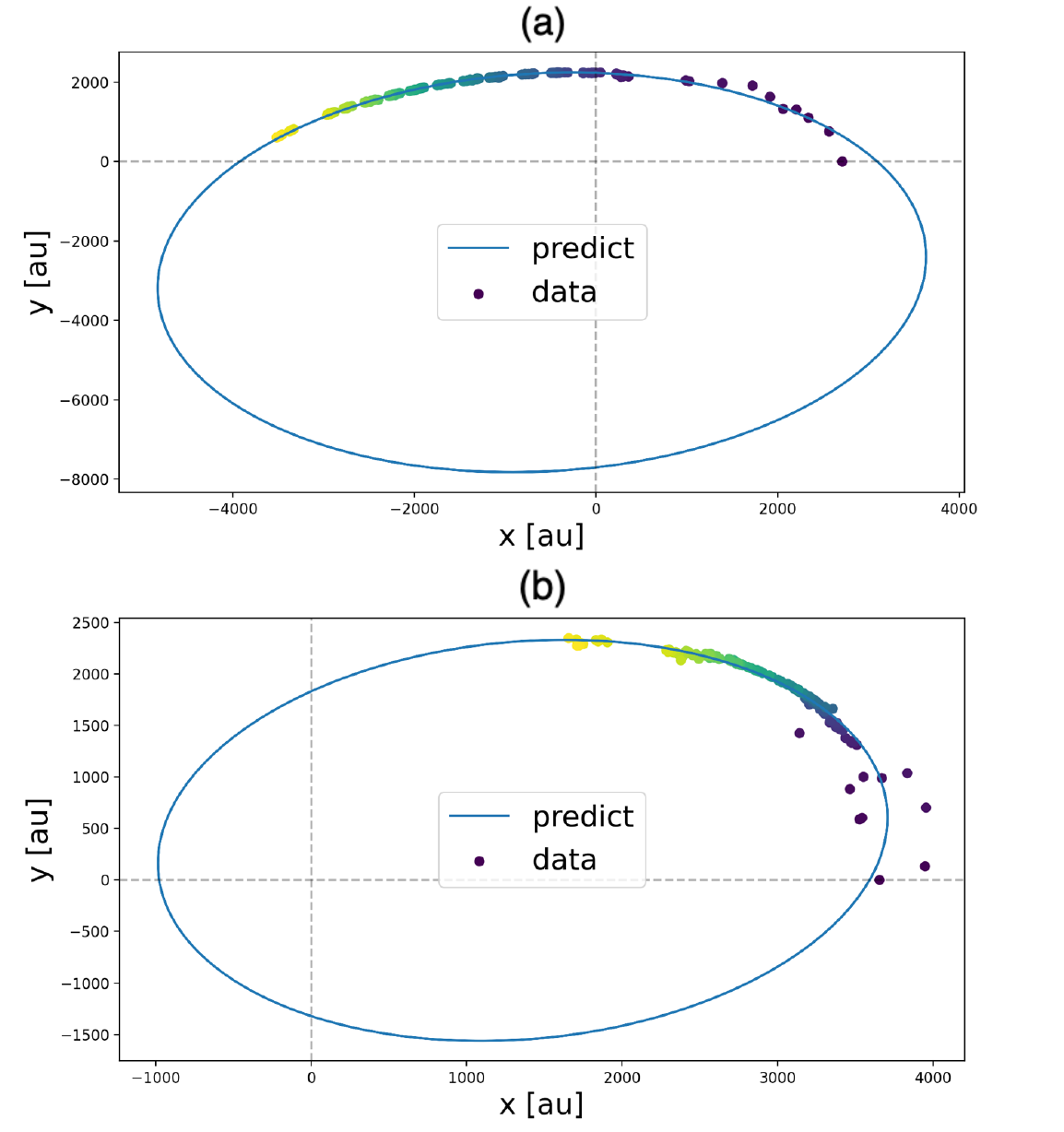}
\caption{Predicted orbital trajectories for (a) S1 and (b) S9 obtained using the optimal $w$ values identified in Fig.~\ref{f7}. Colored dots represent the observed astrometric data, with colors indicating orbital phase.}
\label{f8}
\end{figure}

Figure~\ref{f8} shows the corresponding predicted orbital trajectories for (a) S1 and (b) S9. Note that weight parameter of $w = 10^{-3}$ is adopted for both models. As in the case of S2, the model achieves smooth and physically plausible trajectories, trained on the available astrometric data. Despite the restricted orbital coverage, the PINNs capture the overall orbital shapes with reasonable fidelity.

Of particular interest is the fact that both S1 and S9 are accompanied by a relatively large number of observational data points. However, due to their long orbital periods compared to S2, the observed arcs span only limited segments of their full orbits. While S2 completes a full revolution during the observational period, S1 and S9 cover only partial arcs, limiting the completeness of their training datasets.

\begin{table}
    \centering
    \caption{Fundamental parameters and outcomes of our analysis for S1, S2, and S9, listing their orbital periods, data coverage, relativistic precession angles, and the inferred upper limits on the cosmological constant $\Lambda$}
    \begin{tabular}{|c|c|c|c|c|}\hline
    & Period [Yr] & Data count & $\delta\phi_{\rm SP}$ ['] & $\Lambda$ $[m^{-2}]$ \\ \hline
    S1 & 166 & 161 & 0.79 & $2.42\times10^{-41}$ \\ \hline
    S2 & 16 & 145 & 12.15 & $5.67\times10^{-40}$ \\ \hline
    S9 & 51 & 160 & 1.78 & $9.84\times10^{-41}$ \\ \hline
    \end{tabular}
    \label{table2}
\end{table}

The key orbital and observational parameters for the three stars analyzed—S1, S2, and S9—are summarized in Table~\ref{table2}, including their orbital periods, number of data points, Schwarzschild precession angles, and the resulting upper bounds on $\Lambda$. The values of $\delta\phi_{\rm SP}$ and $\Lambda$ for S1 and S9 were obtained following the same procedure used for S2 (see \S~4.2). Remarkably, both S1 and S9 provide upper bounds that are an order of magnitude tighter than the S2-based constraint.

In principle, full coverage of the orbital phase improves the robustness of regression and dynamical inference. Nonetheless, the results for S1 and S9 exhibit apparently tighter constraints on $\Lambda$ than those from S2. This counterintuitive result arises because Schwarzschild precession, which dominates relativistic effects near the central mass, weakens for stars with longer orbital periods. Consequently, the relative contribution of cosmological precession becomes more prominent, allowing for clearer separation between the total precession $\delta\phi_{\rm Reg}$ and the Schwarzschild term $\delta\phi_{\rm SP}$.

\begin{figure}
\begin{center}
    \includegraphics[scale=0.35]{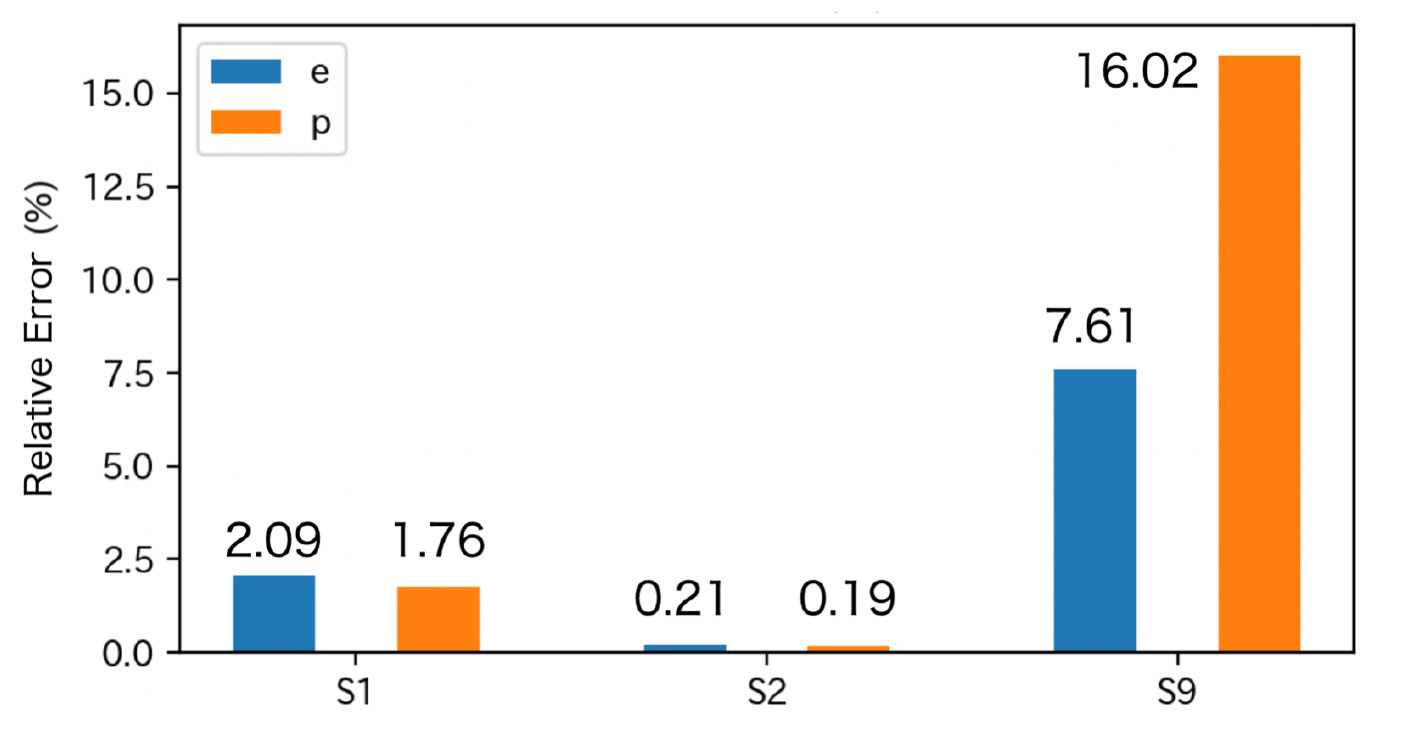}
    \caption{Relative error ($\%$) in eccentricity $e$ (blue) and semi-latus rectum $p$ (orange) for each star. While S2 shows sub-percent-level precision, S1 and S9 exhibit substantially larger uncertainties.}
    \label{f9}
\end{center}
\end{figure}

However, a careful comparison of the inferred orbital parameters and their associated uncertainties reveals a critical caveat. Figure~\ref{f9} presents the relative errors in the predicted eccentricity $e$ and semi-latus rectum $p$ for each star, computed as:
\begin{equation}
\mathmakebox[0.9\columnwidth]{
{\rm Error}=\frac{|z_{\rm pred}-z_{\rm data}|}{z_{\rm data}} \;,
}
\end{equation}
where $z = e$ or $p$. Although S2 yields a sub-percent accuracy, the errors in S1 and S9 are one to two orders of magnitude larger, particularly for $p$ in S9, where the relative error exceeds 16\%.

This discrepancy is attributed to the limited phase coverage of S1 and S9. With only partial orbits available, the PINNs are more prone to overfitting local structures, including observational noise and sampling artifacts. This leads to distorted inference of global orbital parameters, affecting the accuracy of precession angles and thus undermining the reliability of the resulting constraints on $\Lambda$.

We therefore emphasize that the apparent tightness of a constraint should not be interpreted in isolation. Reliable inference demands not only an optimal balance between physical and observational losses, but also comprehensive and high-quality data. Taking these factors into account, we conclude that the S2-based estimate provides the most robust and trustworthy upper bound on the cosmological constant among the cases analyzed.

This quantitative comparison highlights the importance of not only obtaining strong constraints, but also rigorously assessing their robustness through error diagnostics. These findings further suggest that long-period stars are not necessarily better suited for constraining $\Lambda$, unless their orbits are sufficiently sampled with high phase coverage.

\section{Summary \& Future Perspectives}
In this study, we proposed a novel analytical framework using Physics-Informed Neural Networks (PINNs) to constrain the cosmological constant $\Lambda$ by analyzing the relativistic precession of stellar orbits around the SMBH Sgr A* at the Galactic center. In particular, we focused on the S2 star, whose complete orbital revolution within the observational period provides a robust data set for dynamical inference.

By comparing the total precession inferred from the astrometric data with the theoretically predicted Schwarzschild precession, we isolated the contribution from the cosmological constant. This yielded a stringent upper bound of $\Lambda \leq 5.67 \times 10^{-40}, \mathrm{m}^{-2}$—approximately two orders of magnitude tighter than previous estimates \citep{galikyan23}.

The results demonstrate the utility of PINNs in reconstructing orbital dynamics from noisy and incomplete data, by incorporating governing physical laws directly into the learning process. The use of inverse PINNs (iPINNs) further enabled us to extract orbital elements directly from data, improving precision and physical consistency.

To test the generality of our method, we extended the analysis to two additional long-period S-stars, S1 and S9. For these stars, the relativistic Schwarzschild precession is intrinsically weaker due to their extended orbital timescales, making the contribution from $\Lambda$ comparatively more prominent. Indeed, the derived upper bounds on $\Lambda$ from S1 and S9 were numerically tighter than that from S2.

However, due to the limited phase coverage of S1 and S9—each tracing only a partial arc of their full orbits—the iPINN-based inference exhibited significantly larger uncertainties in the derived orbital parameters. This suggests potential overfitting to localized data features and highlights the critical importance of high-quality and phase-complete observational data for reliable dynamical inference.

In conclusion, although long-period stars may in principle amplify the relative effect of cosmological precession, their utility is strongly conditioned by the completeness and precision of the available data. Among the cases analyzed, the S2 star—thanks to its well-sampled full orbit—provides the most reliable constraint. We thus adopt the S2-based upper bound as our final estimate for the cosmological constant.

The results obtained in this study open promising directions for future work at the intersection of machine learning and astrophysics. In particular, upcoming high-precision observations from next-generation facilities such as the ELT (Extremely Large Telescope) \citep[e.g.,][]{epchtein+07, paumard+10} and ongoing campaigns with the VLT will be instrumental in detecting subtle orbital shifts, enabling deeper probes of relativistic and cosmological effects.
These advancements are expected to further refine constraints on the cosmological constant and potentially determine the spin of Sgr A$^\star$ \citep{yao+25}.

Enhanced astrometric accuracy will allow more detailed measurements of precession anomalies, improving not only the precision of $\Lambda$ constraints but also the ability to test alternative gravitational theories on galactic scales \citep[e.g.,][]{galikyan24,tan+24}. 
The iPINN-based framework developed here is broadly applicable to other stellar systems, both within and beyond the Galaxy, offering a scalable tool for dynamical inference and cosmological diagnostics.

Moreover, future advances in machine learning—particularly improved strategies for balancing physical and regression losses, and the incorporation of adaptive training algorithms—are expected to further increase the robustness of parameter estimation, even in the presence of noise or missing data.

In this context, the integration of physical models with data-driven approaches, as demonstrated in this study, offers a powerful paradigm for tackling complex problems in astrophysics. As both observational capabilities and computational techniques continue to advance, our approach is poised to become a key methodology for exploring the large-scale structure and dynamics of the Universe.

\section*{Acknowledgements}
This work was supported by the University grant No.GR2302 of Fukuoka University, and also by JSPS KAKENHI Grant Number (JP18K03700, JP21K03612, JP23K25895, JP24K00631 and JP25K07374). This research was also supported by MEXT as “Program for Promoting researches on the Supercomputer Fugaku” (Structure and Evolution of the Universe Unraveled by Fusion of Simulation and AI; Grant Number JPMXP1020230406) and JICFuS.

\section*{DATA Availability}
The data underlying this article will be shared on reasonable request to the corresponding author.

\bibliographystyle{mnras}
\bibliography{example}








\bsp	
\label{lastpage}
\end{document}